\documentclass[aps,twocolumn,superscriptaddress,altaffilletter,lengthcheck,
tightenlines,showpacs,showkeys]{revtex4}
\usepackage{multirow}
\newcommand{\al}{\alpha}

\newcommand{\ben}{\begin{eqnarray}}
\newcommand{\een}{\end{eqnarray}}
\newcommand{\be}{\begin{equation}}
\newcommand{\ee}{\end{equation}}
\newcommand{\ba}{\begin{eqnarray}}
\newcommand{\ea}{\end{eqnarray}}
\newcommand{\n}{\label}

\newcommand{\ga}{\gamma}
\newcommand{\ro}{\rho}

\newcommand{\bn}{\begin{equation}\label}

\usepackage[dvipdf]{epsfig}
\usepackage{color}
\begin{document}
\title{Interacting dark sector with variable vacuum energy
\\  }
\author{Luis P. Chimento}
\affiliation{Departamento de F\'{\i}sica, Facultad de Ciencias Exactas y Naturales,  Universidad de Buenos Aires and IFIBA, CONICET, Ciudad Universitaria, Pabell\'on I, Buenos Aires 1428 , Argentina}
\author{Mart\'{\i}n G. Richarte}
\affiliation{Departamento de F\'{\i}sica, Facultad de Ciencias Exactas y Naturales,  Universidad de Buenos Aires and IFIBA, CONICET, Ciudad Universitaria, Pabell\'on I, Buenos Aires 1428 , Argentina}
\author{Iv\'an E. S\'anchez Garc\'ia.}
\affiliation{Departamento de F\'{\i}sica, Facultad de Ciencias Exactas y Naturales,  Universidad de Buenos Aires and IFIBA, CONICET, Ciudad Universitaria, Pabell\'on I, Buenos Aires 1428 , Argentina}
\date{\today}
\bibliographystyle{plain}
\begin{abstract}
We examine a cosmological scenario where dark matter is coupled to a variable vacuum energy  while baryons and photons  are two decoupled components   for a  spatially flat Friedmann-Robertson-Walker spacetime.  We apply the $\chi^{2}$ method to the updated observational Hubble data for constraining the cosmological parameters and  analyze the amount of dark energy in the radiation era. We show that our model fulfills the severe bound of $\Omega_{x}(z\simeq 1100)<0.009$ at  the $2\sigma$ level, so it is consistent with the recent analysis that includes cosmic microwave background anisotropy measurements from the Planck survey, the Atacama Cosmology Telescope, and  the South Pole Telescope along with the future constraints achievable by the Euclid and CMBPol experiments, and fulfills the stringent bound  $\Omega_{x}(z\simeq 10^{10})<0.04$ at the $2\sigma$ level in the big-bang nucleosynthesis epoch.
\end{abstract} 
\vskip 1cm
\bibliographystyle{plain}
\maketitle
\section{Introduction}
The existence of dark matter and dark energy have been supported by many observations  such as  cosmic microwave background, power spectrum  of clustered matter, and  supernovae Ia data  among other probes \cite{Book, Planck2013, WMAP9}.  Even though a fundamental (microscopic) theory for describing the dark sector remains elusive, observations suggest that 
the dark matter  is connected with the formation of the large-structure \cite{Book} whereas   dark energy  is causing the expansion of the Universe to speed up, rather than slow down. 

To understand the hidden nature of the interacting dark sector \cite{jefe1}, one has to confront these models with the observational data; the transfer of energy could alter the cosmic history  leading to testable imprints in the Universe \cite{CR1}. The amount of dark energy in the recombination epoch should fulfill the bound by $\Omega_{\rm x}(z\simeq 1100)<0.01$ at least \cite{robbers}.  Additionally, the presence of vacuum energy during big-bang nucleosynthesis (BBN)  is well motivated
both by considerations of dark energy as well as inflation, giving as a stringent bound $\Omega_{\rm x}({\rm 1Mev}) < 0.21$ so that the model does not affect the BBN process and therefore the abundance of light elements such as $^{4}{\rm He}$ \cite{KT}, \cite{KT2}. The Planck mission indicates that $\Omega_{\rm x}(z\simeq 1100)<0.009$ with a $95\%$ C.L., whereas  the joint analysis based on  future surveys (Euclid+CMBPol)  will be able to constrain 
$\Omega_{\rm x}(z\simeq 1100)$ down to  $0.00092$; the joint analysis of the Euclid+Planck data will be  less restrictive because it will lead to  $\Omega_{\rm x}(z\simeq 1100) <0.0022$ \cite{Euc}. 

Our goal is to investigate a universe  with an interacting dark sector and two decoupled components.  We constrain the cosmic set of parameters by using the updated Hubble data and  the severe bounds reported by the Planck mission on early dark energy among other observational data.

\section{The model}

We consider a spatially flat homogeneous  and isotropic universe described by  Friedmann-Robertson-Walker (FRW) spacetime with a line element given by  $ds^{2}=-dt^{2}+a^{2}(t) (dx^{2}+dy^{2}+dz^{2})$, with  $a(t)$ being  the scale factor. The universe is  filled with  interacting dark matter  and variable vacuum energy (VVE) plus decoupled baryonic matter and radiation components.    The evolution of the FRW universe is governed by  the Friedmann and conservation equations,   
\be
\n{01}
3H^{2}=\ro_{\rm t}=\ro_{\rm r}+ \ro_{\rm b}+\ro_{\rm m}+ \ro_{\rm x},
\ee
\bn{rr}
\dot\ro_{\rm r}+3H\ga_{\rm r}\ro_{\rm r}=0,~~~\dot\ro_{\rm b}+3H\ga_{\rm b}\ro_{\rm b}=0,
\ee
\be
\n{02}
\dot{\ro}_{\rm m}+\dot{\ro}_{\rm x}+3H(\ga_{\rm m}\rho_{\rm m}+\ga_{\rm x}\rho_{\rm x})=0,
\ee
where $H = \dot a/a$ is the Hubble expansion rate, and the equations of state  for each species take a barotropic form $p_{\rm i}=(\gamma_{\rm i}-1)\ro_{\rm i}$. Then the constants $\ga_{\rm i}$ indicate the barotropic index of each component being ${\rm i}=\{ \rm r,b,m,x\}$, so  that  $\gamma_{\rm r}=4/3$, $\ga_{\rm b}=1$, and $\gamma_{\rm x}=0$,  whereas  $\ga_{\rm m}$ will be estimated later on. So, $\rho_{\rm x}$ plays the role of a VVE, $\ro_{\rm b}$ represents a pressureless baryonic matter,  $\ro_{\rm r}$ is a radiation  component and $\rho_{\rm m}$ can be associated with   dark matter. 

Solving the linear algebraic system of Eq. (\ref{02}) along with $\ro=\ro_{\rm m}+\ro_{\rm x}$, we acquire both dark component densities as functions of $\ro$ and $\ro'$
\be
\n{04}
\ro_{\rm m}=  - \frac{\ga_{\rm x} \ro +\ro '}{ \ga_{\rm m}-\ga_{\rm x}}, \qquad \ro_x= \frac{\ga_{\rm m} \ro +\ro '}{ \ga_{\rm m}-\ga_{\rm x}},
\ee
where we introduced the variable $\eta = \ln(a/a_0)^{3}$, $' \equiv d/d\eta$.  The baryons and photons are decoupled from the dark sector, so Eq. (\ref{rr}) leads to  $\ro_{\rm r}=\ro_{\rm r0}a^{-3\ga_{\rm r}}$ and
$\ro_{\rm b}=\ro_{\rm b0}a^{-3\ga_{\rm b}}$, respectively. 
In order to continue the analysis of the interacting dark sector, we  take into account the exchange of energy in the dark sector through the term $3HQ$ into 
 Eq. (\ref{02}), which we will tackle next. It is convenient to use $d\eta=3Hdt$  to obtain the  balance equations in a  simpler form:
\be
\label{03}
{\ro'}_{\rm m}+\ga_{\rm m}\rho_{\rm m}=-Q,  \qquad {\ro'}_{\rm x}+\ga_{\rm x}\rho_{\rm x}=Q. 
\ee
From Eqs. (\ref{04}) and (\ref{03}), we obtain the source equation \cite{jefe1} for the energy density $\ro$ of the dark sector
\be
\n{14}
\ro''+(\ga_{\rm m}+ \ga_{\rm x})\ro' + \ga_{\rm m}\ga_{\rm x}\ro =  Q( \ga_{\rm m}-\ga_{\rm x}).
\ee
Here, the nonlinear interaction $Q$ between both dark components is  $Q=\alpha\ro'\ro$, with $\alpha$ being  the coupling constant. This  interaction was not examined in the literature before  and  gives rise to a scenario where  VVE can be viewed as a variable cosmological constant \cite{LV}. 

By replacing the specific form of $Q$ into the source equation (\ref{14}),  it turns into a nonlinear second-order differential equation for the total energy density $\ro$. Inserting $\ga_{\rm x}=0$ into the latter equation, one gets its first integral
\be
\label{IP}
\ro'=\ga_{\rm m}\left[\frac{\al}{2}\ro^2-\ro+ {\cal D}\right],
\ee
where $ {\cal D}$ is an integration constant. Plugging Eq. (\ref{IP}) into Eq. (\ref{04}), we obtain that $\ro_{\rm x}=[\al\ro^2/2+ {\cal D}]=\Lambda$, so  dark energy can be considered as a VVE provided at late times $\ro_{t}\simeq \ro=3H^{2}$ and then $\Lambda\simeq (\al H^4/2+ {\cal D}) $ \cite{LV}. In order to get $\rho(a)$, we need to express the  first-order nonlinear differential equation  (\ref{IP}) as an integration by quadrature. Solving Eq. (\ref{IP})  under the condition $1>2\al  {\cal D}$ allows us to obtain the total energy density of the dark sector 
\be
\label{rod14}
\ro=\frac{{\cal K}(1+{\cal R})a^{-3\ga_{\rm m} {\cal R}}+{\cal R}-1}{\alpha \left[{\cal K}a^{-3\ga_{\rm m} {\cal R}}-1\right]},
\ee
where ${\cal R}=\sqrt{1-2\alpha {\cal D}}$, ${\cal K}$ is an integration constant.  Using  the present-density parameters $\Omega_{i0}=\ro_{i0}/3H_{0}^2$ and the flatness condition, $1=\Omega_{\rm r0}+\Omega_{\rm b0}+\Omega_{\rm x0}+\Omega_{\rm m0}$, we write the integration constants   ${\cal K}$ and  ${\cal D}$ in terms of density parameters:
\[{\cal K}=\frac{3\al{H_0}^2(\Omega_{\rm x0}+\Omega_{\rm m0})-(1-{\cal R})}{3\al{H_0}^2(\Omega_{\rm x0}+\Omega_{\rm m0})-(1+{\cal R})}, \] 
\be
\label{K}
{\cal D}=3{H_0}^2\Omega_{\rm x0}-\frac{\alpha}{2}[3{H_0}^2(\Omega_{\rm x0}+\Omega_{\rm m0})]^2.
\ee
The total energy density  is given by
\[\ro_{\rm t}= 3{H_0}^2(1-\Omega_{\rm b0}-\Omega_{\rm x0}-\Omega_{\rm m0})a^{-3\ga_{\rm r}}+ 3{H_0}^2\Omega_{\rm b0}a^{-3} \]
\be
\n{ET}
+ \frac{{\cal K}(1+{\cal R})a^{-3\ga_{\rm m} {\cal R}}+{\cal R}-1}{\alpha \left[{\cal K}a^{-3\ga_{\rm m} {\cal R}}-1\right]}.
\ee
The Universe is dominated by radiation  at early times. After this epoch  pressureless baryonic matter dominates  followed by  an era dominated by  dark matter when $\ga_{\rm m} {\cal R}\simeq 1$, ending with a de Sitter phase at late times [cf. Eq. (\ref{ET})].  To see that  dark matter dominates the evolution of the Universe  during a short period of time, we  use energy density of the dark sector (\ref{rod14}) and  we find  $\ro=(1/\al)-({\cal R}/\al)[1+x/(1-x)]$, where $x={\cal K}a^{-3\ga_{\rm m} {\cal R}}$. When $x$ is considerably small,  Eq. (\ref{rod14}) leads to $\ro\simeq (1-{\cal R})/\al-(2{\cal R}/\al)x$  along with $\ro_{\rm m}=-\ro'/\ga_{\rm m}\simeq (2{\cal R}/\al\ga_{\rm m})x'$. Using $x'=-{\cal K}\ga_{\rm m} {\cal R}a^{-3\ga_{\rm m} {\cal R}}$, we obtain that  $\ro_{\rm m}\simeq -(2{\cal R}^2{\cal K}/\al\ga_{\rm m})a^{-3\ga_{\rm m} {\cal R}}>0$ provided ${\cal K}<0$; such fact can be verified  by using the best-fit values of the cosmological parameters found in the next section.
\section{Observational Hubble data constraints}
We focus on the cosmological constraints on the parameter space of the interacting dark sector  plus the decoupled radiation and baryonic  components. The  statistical analysis is based on the $\chi^{2}$ function of the Hubble data which is constructed as (see,  e.g., Ref. \cite{Press})
\be
\n{c1}
\chi^2(\theta) =\sum_{k=1}^{29}\frac{[H(\theta,z_k) - H_{\rm obs}(z_k)]^2}{\sigma(z_k)^2},
\ee
where the $\theta$ denotes a set of parameters, $H_{\rm obs}(z_k)$ is the updated observational $H(z)$ data at the redshift $z_k$,  and $\sigma(z_k)$ is the corresponding $1\sigma$ uncertainty \cite{Hdata}. We will use  data listed in  Ref. \cite{Hdata} and the Hubble parameter at $z=0$ \cite{Riess}, commenting how the observational data  were obtained \cite{Hdata}. The  first compilation of  nine  measurements of the Hubble parameter as a function of the redshift
in the range $0.1 < z < 1.75$ was used for constraining a dark energy model in Ref. \cite{Verdea}. 
Stern \emph{et al.}   added two more observations in the redshift range $0.2 < z < 1.0$ \cite{Verdeb}, and 
another eight new high-accuracy estimates of $H(z)$
led to a sample of $19$ observational 
$H(z)$ measurements, spanning almost 10 Gyr of cosmic time \cite{Verdec}. Further, 
Blake \emph{et al}. obtained three  data points by  combining 
a baryon acoustic peak and an Alcock-Paczynski distortion from galaxy clustering 
in the WiggleZ Dark Energy Survey \cite{Verded}, whereas from the
Sloan Digital Sky Survey Data Release 7 within a redshift window $ 0< z<0.4$, Zhang \emph{et al.} obtained
four new observational $H(z)$ data  points \cite{Verdee}.  Another two data points  were added by considering  the baryonic acoustic oscillations scale as a standard ruler in the radial direction \cite{Verdex}.    Using  Eqs. (\ref{K}) and (\ref{ET}), one gets the Hubble parameter in terms of the redshift, $z=a^{-1}-1$, and the relevant cosmological parameters  $\theta=(H_0,\Omega_{\rm b0},\Omega_{\rm x0},\Omega_{\rm m0},\alpha,\gamma_{\rm m})$ as follows,  $H(z, \theta)=[\ro_{\rm t}/3]^{1/2}$. 
  
\begin{figure}[hbt!]
\begin{minipage}{1\linewidth}
\resizebox{1.4in}{!}{\includegraphics{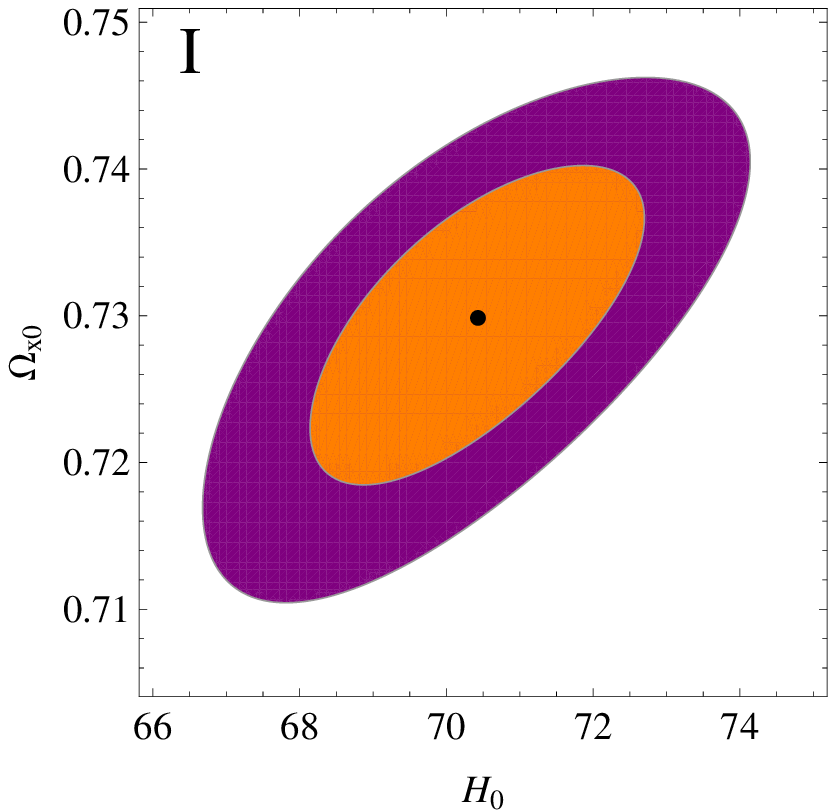}}\hskip0.06cm
\resizebox{1.4in}{!}{\includegraphics{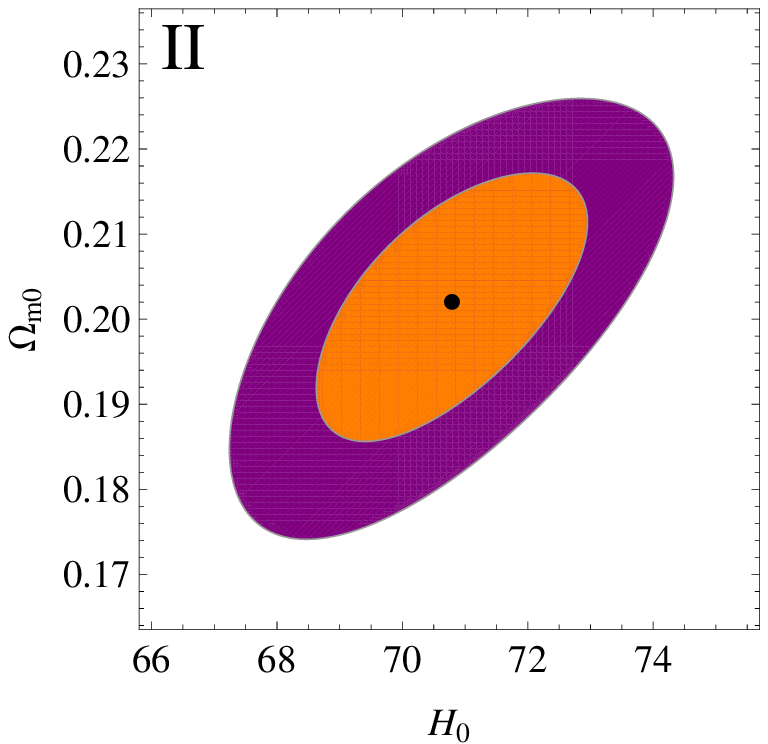}}\hskip0.06cm
\resizebox{1.4in}{!}{\includegraphics{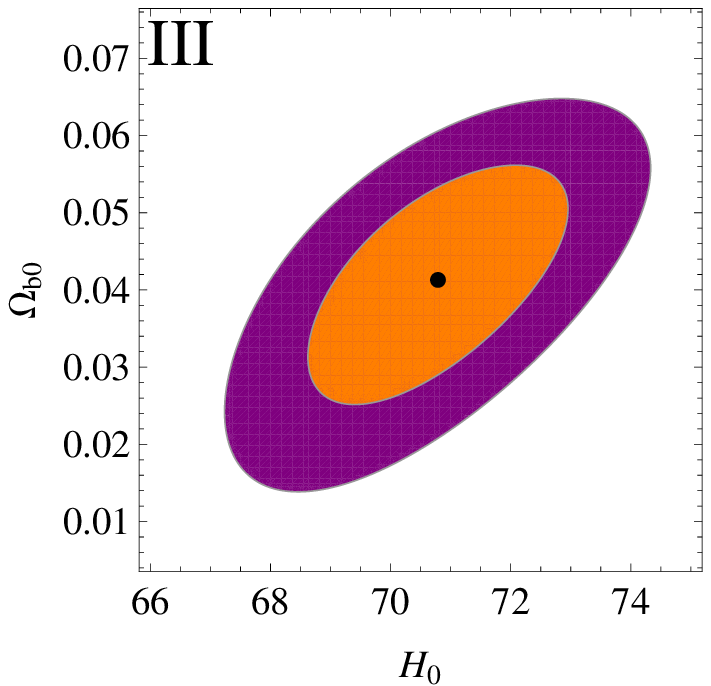}}\hskip0.06cm
\resizebox{1.4in}{!}{\includegraphics{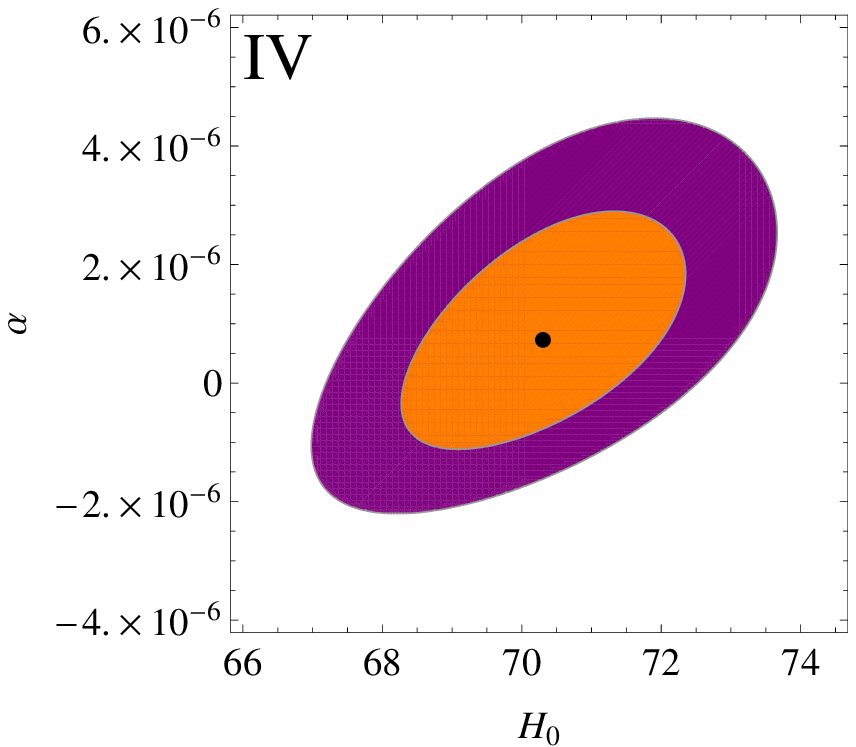}}\hskip0.06cm
\resizebox{1.4in}{!}{\includegraphics{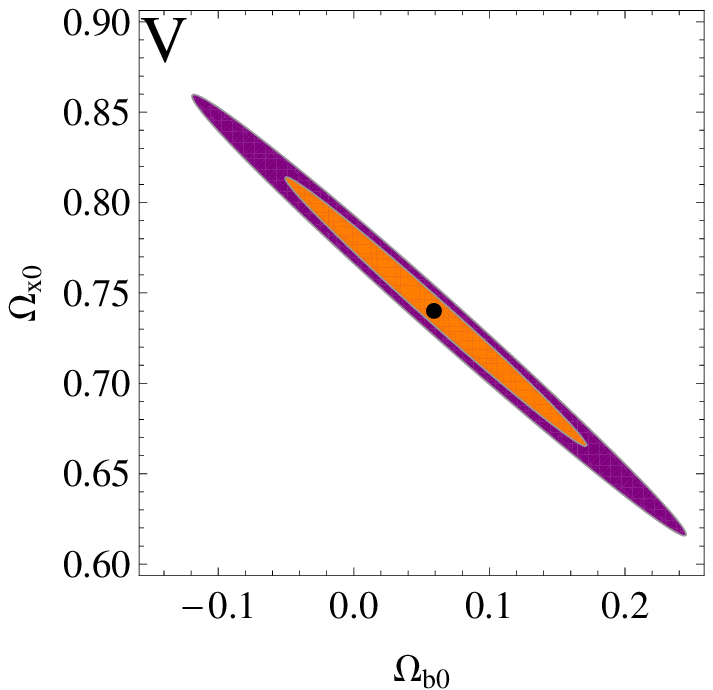}}\hskip0.06cm
\resizebox{1.4in}{!}{\includegraphics{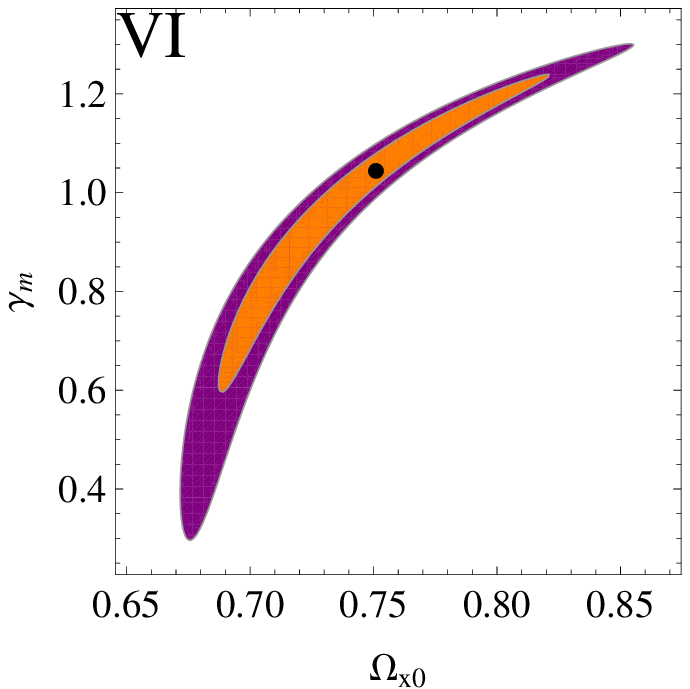}}\hskip0.06cm
\caption{\scriptsize{Two-dimensional C.L. associated with $1\sigma$, $2\sigma$ for different $\theta$ planes.}}
\label{F1}
\end{minipage}
\end{figure}

\begin{table}[ht!]
\centering
\scalebox{0.58}{
\begin{tabular}{|l|l|l|l|}
  \hline
  \multicolumn{4}{|c|}{$2D$ ${\rm Confidence~~ level}$} \\
  \hline
  ${\rm N}$ & ${\rm Priors}$ & ${\rm Best fits}$& $\chi^2_{\rm d.o.f}$ \\
  \hline
  I & $(\Omega_{\rm b0}, \Omega_{\rm m0}, \alpha,  \ga_{\rm m})=(0.051, 0.219, 10^{-6},  1.010)$ & $(H_{0}, \Omega_{\rm x0})=(70.40^{+2.31}_{-2.24}, 0.73\pm 0.01)$ & $0.725$ \\
  \hline
  II & $(\Omega_{\rm b0}, \Omega_{\rm x0}, \alpha,  \ga_{\rm m})=(0.049, 0.74, 10^{-7},  1.014)$ & $(H_{0}, \Omega_{\rm m0})=(70.79^{+2.11}_{-2.19}, 0.202^{+0.015}_{-0.016})$ & $0.703$ \\
  \hline
  III & $(\Omega_{\rm x0}, \Omega_{\rm m0}, \alpha,  \ga_{\rm m})=(0.74, 0.21, 10^{-7}, 1.014)$ & $(H_{0}, \Omega_{\rm b0})=(70.79^{+2.18}_{-2.17}, 0.041^{+0.015}_{-0.016})$ & $0.703$ \\
  \hline
  IV & $(\Omega_{\rm b0}, \Omega_{\rm x0}, \Omega_{\rm m0},  \ga_{\rm m})=(0.049, 0.73, 0.220,  1.010)$ & $(H_{0}, \alpha)=(70.30^{+2.05}_{-2.04}, [7.5^{+7.21}_{-7.61}]\times 10^{-7})$ & $0.724$ \\
  \hline
  V & $(H_{0}, \Omega_{\rm m0}, \alpha, \ga_{\rm m})=(69.04, 0.20, 10^{-7},  1.036)$ & $(\Omega_{\rm b0}, \Omega_{\rm x0})=( 0.059^{+0.113}_{-0.107}, 0.74^{+0.07}_{-0.07})$ & $0.753$ \\
  \hline
  VI & $(H_{0}, \Omega_{\rm b0}, \Omega_{\rm m0}, \alpha)=(68.5, 0.043, 0.205, 10^{-8})$ & $(\Omega_{\rm x0}, \ga_{\rm m0})=( 0.77^{+0.07}_{-0.06}, 1.047^{+0.192}_{-0.452})$ & $0.794$ \\
  \hline
\end{tabular}}
\caption{\label{I} We show the observational bounds for the 2D C.L. obtained in Fig. (\ref{F1}) by varying two cosmological parameters.}
\end{table}

\begin{figure}[hbt!]
\begin{minipage}{1\linewidth}
\resizebox{2.4in}{!}{\includegraphics{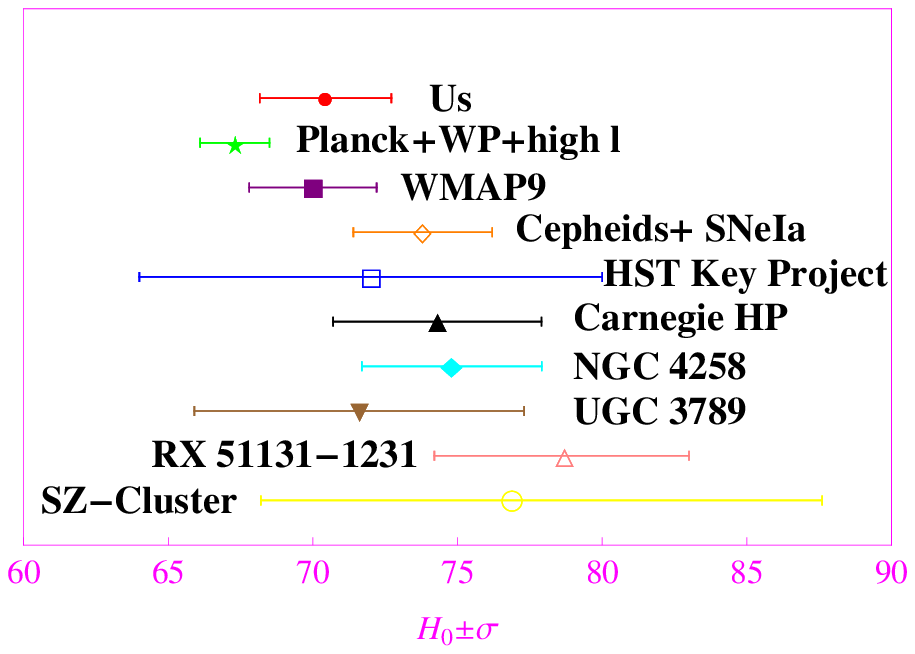}}\hskip0.06cm
\resizebox{2.4in}{!}{\includegraphics{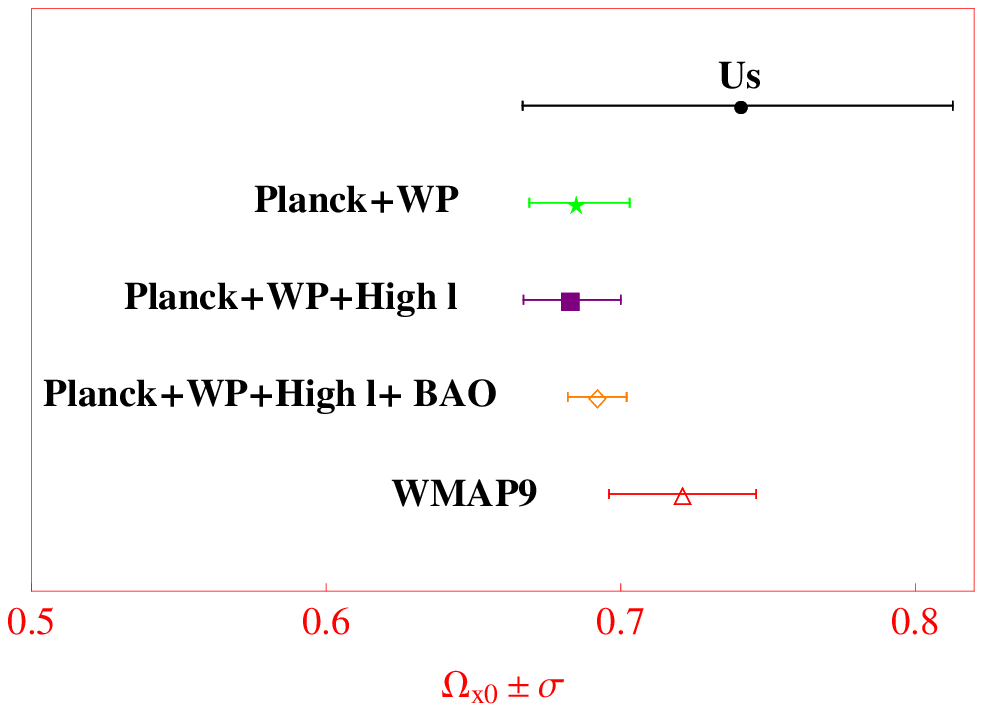}}\hskip0.06cm
\resizebox{2.4in}{!}{\includegraphics{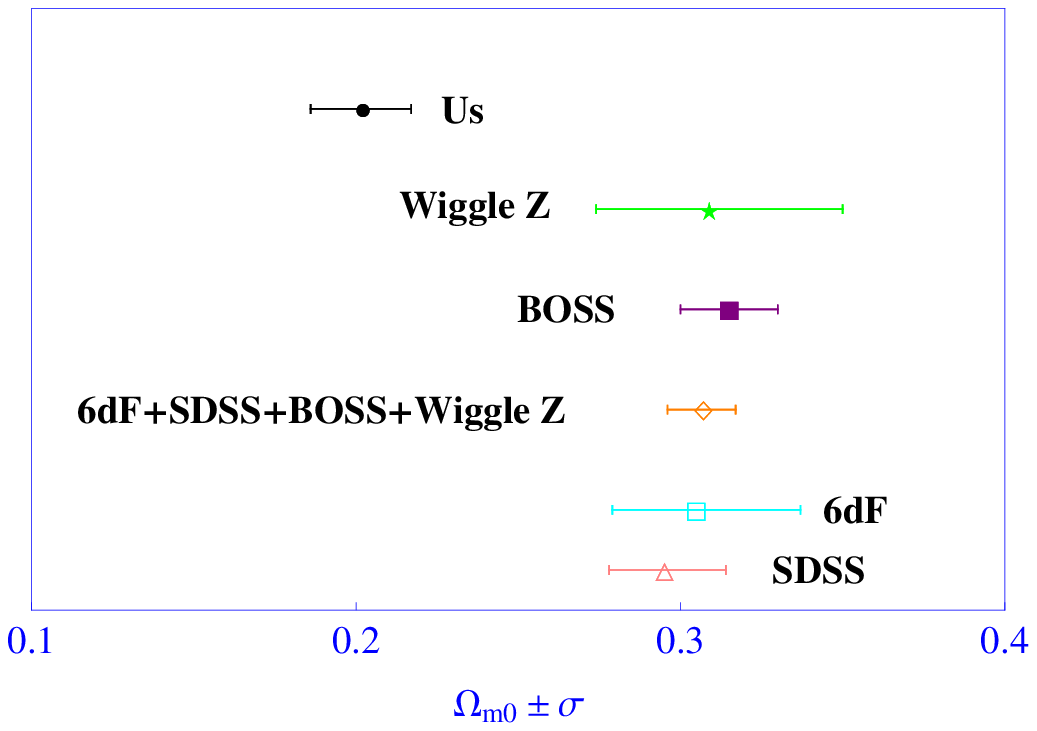}}\hskip0.06cm
\caption{\scriptsize{Comparison of Hubble parameter, dark energy  and dark matter amounts
, with estimates of
$±\sigma$ errors, from a number of different methods.}}
\label{F2}
\end{minipage}
\end{figure}

\begin{figure}[hbt!]
\begin{minipage}{1\linewidth}
\resizebox{2.4in}{!}{\includegraphics{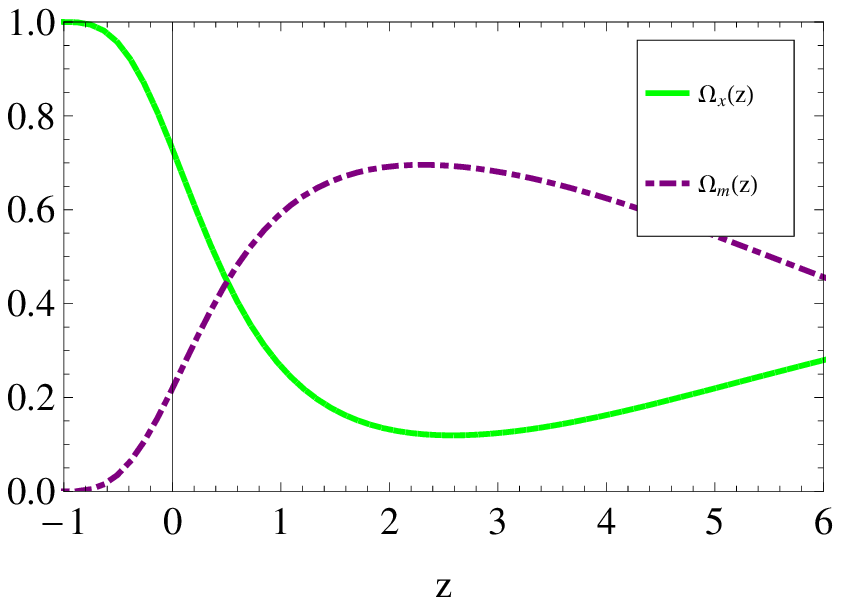}}\hskip0.06cm
\resizebox{2.4in}{!}{\includegraphics{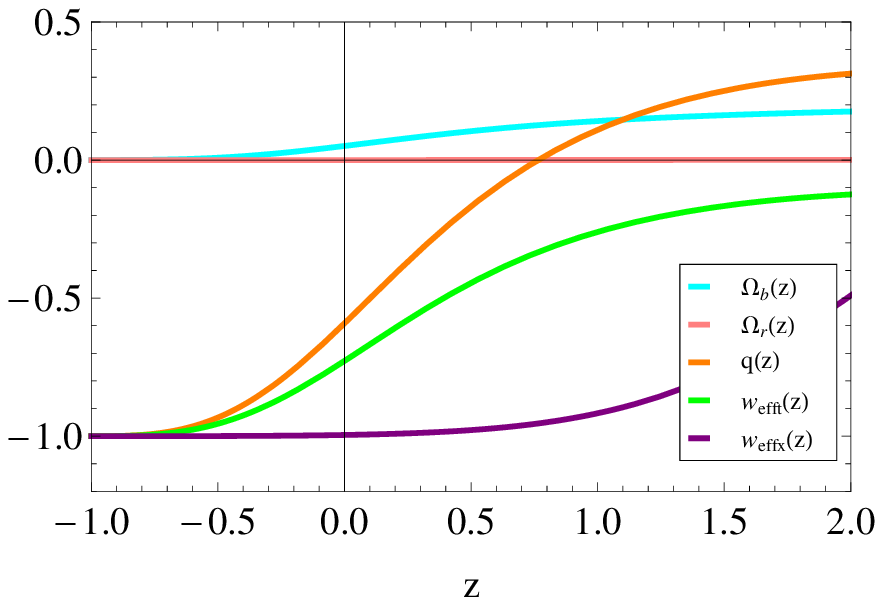}}\hskip0.06cm
\caption{\scriptsize{Upper panel: Plot of $\Omega_{\rm x}$ and  $\Omega_{\rm m}$ in terms of  the redshift $z$. Lower panel: Plot $\Omega_{\rm b}$,  $\Omega_{\rm r}$,  ${\rm w}_{\rm eff t}$, ${\rm w}_{\rm eff x}$, and $q$  in terms of  the redshift $z$. }}
\label{F3}
\end{minipage}
\end{figure}

\begin{table}[ht!]
 \scalebox{0.8}{
\centering 
\begin{tabular}{|c|c|r|r|r|r|r|r|r|r|r|r|r|r|}
	\hline
Mission &  {\rm Bound~ on~ } $ \Omega_{x}[z\simeq 10^{3})]$\\
	\hline	
{\rm Us} &  $\leq 0.000001$ \\
{\rm Euclid} &  $<0.024$ \\
{\rm CMBPol} &  $<0.0025$ \\
{\rm SPT} &  $<0.02$ \\
{\rm WMAP7} &  $<0.018$ \\
{\rm WMAP7+ACT} &  $<0.025$ \\
{\rm WMAP7+SPT} &  $<0.013$ \\
{\rm WMAP7+SPT+BAO+SNe} &  $<0.014$ \\
{\rm Planck+WP} &  $<0.010$ \\
{\rm Planck+WP+high~L } &  $<0.009$ \\
{\rm Planck+Euclid} &  $<0.0022$ \\
{CMBPol+Euclid} &  $<0.00092$ \\
\hline
\end{tabular}}
\caption{Comparison of different estimations or simulations on the fraction of dark energy at early times.} 
\label{II} 
\end{table}

The two-dimensional C.L. obtained with a standard $\chi^{2}$ function for two parameters is shown in Fig. (\ref{F1}), and the estimation of these parameters is briefly summarized in Table (\ref{I});  reporting  their  corresponding marginal  $1\sigma$ error bars \cite{Bayes}.  We find  the best fit at  $(\Omega_{\rm b0}, \Omega_{\rm x0})=( 0.059^{+0.113}_{-0.107}, 0.74^{+0.07}_{-0.07})$ with  $\chi^2_{\rm d.o.f}= 0.753$ by using the priors $(H_{0}, \Omega_{\rm m0}, \alpha, \ga_{\rm m})=(69.04{\rm km~s^{-1}\,Mpc^{-1}}, 0.20, 10^{-7},  1.036)$; the values of  $\Omega_{\rm b0}$ and  $\Omega_{\rm x0}$ agree with the data released by the  Planck mission \cite{Planck2013} and  WMAP9 project \cite{WMAP9} [see Fig. (\ref{F2})]. Indeed,   Planck+WP data indicate that  $\Omega_{\rm x0}=0.685 ^{+0.018}_{-0.016}$ at a $68\%$ C.L.;  Planck+WP+high L data lead to  $\Omega_{\rm x0}=0.6830 ^{+0.017}_{-0.016}$ at $68\%$ C.L. \cite{Planck2013} [see Fig. (\ref{F2})]. We  get  the best fit at  $(H_{0}, \Omega_{\rm m0})=(70.79^{+2.11}_{-2.19} {\rm km~s^{-1}\,Mpc^{-1}}, 0.202^{+0.015}_{-0.016})$ along with $\chi^2_{\rm d.o.f}= 0.703<1$. The wiggleZ data give $\Omega_{\rm m0}=0.309^{+0.041}_{-0.035}$ while the joint data 6dF+SDSS+ BOSS+ WiggleZ  lead to $\Omega_{\rm m0}=0.307^{+0.010}_{-0.011}$ at $68\%$ C.L \cite{Planck2013}, showing  a discrepancy on $\Omega_{\rm m0}$ no bigger than  $0.32\%$  [see Figs. (\ref{F2}) and (\ref{F3})].  The analysis  leads to $(\Omega_{\rm x0}, \ga_{\rm m})=( 0.77^{+0.07}_{-0.06}, 1.047^{+0.192}_{-0.452})$, pointing that the dark matter is not pressureless provided the barotropic index is greater than the unity [cf.  Table (\ref{I})]. Regarding  the Hubble parameter, we find that it varies over a wide range, $H_{0} \in [70.30^{+2.05}_{-2.04}; 70.79^{+2.18}_{-2.17} ]{\rm km~s^{-1}\,Mpc^{-1}}$.   Fitting
the base $\Lambda$cold dark matter  model for the WMAP-9 data, it found $ H_{0} =(70.0 \pm  2.2) {\rm km s^{-1} Mpc^{-1}} $ at $68\%$ C.L. \cite{WMAP9}, and agrees with  our best estimations $ H_{0} =70.30^{+2.05}_{-2.04} {\rm km s^{-1} Mpc^{-1}} $ at $68\%$ C.L.  In Fig. (\ref{F2}), we show  bounds of $ H_{0}$ that include the megamaser-based distance to NGC4258, SZ clusters, and others (see  Ref. \cite{Planck2013}).  Figure (\ref{F3}) shows the decelerating parameter, density parameters, and equations of state with the redshift.   Present-day value of $q(z=0) \in [-0.62; -0.59]$ as stated in the  WMAP9 report \cite{WMAP9}. The total equation of state, ${\rm w}_{\rm efft}=-1+\sum_{\rm j}{\ga_{\rm j}\Omega_{\rm j}}$, does not cross the phantom line neither the effective dark energy equation of state, and the same happens for dark energy effective equation of state,  ${\rm w}_{\rm effx}=-[\al\rho\rho'+\rho_{\rm x}]/\rho_{\rm x}$. Their values at $z=0$  vary over the intervals, ${\rm w}_{\rm efft0}\in [-0.74, 0.72]$ and ${\rm w}_{\rm effx0}\in [-0.99, -0.97]$, respectively.

An interacting dark matter--VVE  model has to be constrained with the physics behind
recombination or big-bang nucleosynthesis epochs \cite{CR1}. As is well known, the fraction of dark 
energy in the recombination epoch should fulfill the severe bound $\Omega_{\rm ede}:=\Omega_{\rm x}(z\simeq 1100)<0.01$ \cite{robbers}.  The CMB  measurements will put further  constraints on early dark energy; the latest constraints on early dark energy come from the Planck+WP+high L data: $\Omega_{\rm ede} < 0.009$  at $95\%$ C.L \cite{Planck2013}. We  found that $\Omega_{\rm x}(z\simeq 10^{3}) \in [10^{-6}, 10^{-5}]$, so our estimations satisfied the bound reported by the Planck mission [see Table (\ref{II})]. Further,   the small-scale CMB temperature  measurement from the SPT improves 
over WMAP7 alone by a factor of 3.5 \cite{Reic}, while  WMAP7+SPT+BAO+SNe leads to $\Omega_{\rm ede} < 0.014$, and WMAP+SPT  gives $\Omega_{\rm ede} < 0.013$\cite{Hou}.  Our  value on $\Omega_{\rm x}(z\simeq 1100)\leq 10^{-6}$ at the $1\sigma$ level is  below  the   bounds  achieved  with the forecasting method  applied to the Euclid project \cite{Euc};  this survey will be able to constrain as $\Omega_{\rm ede} < 0.024$ . We fulfill the bound reported from the joint analysis based on  Euclid+CMBPol data,  $\Omega_{\rm ede} <0.00092$ [see Table (\ref{II})].  Our estimation on  $\Omega_{\rm x}(z\simeq 1100)$ is much smaller  than the  bounds obtained  by means of the standard Fisher matrix approach applied to the Euclid and CMBPol experiments \cite{Euc}, \cite{EDE2}.  Around $z=10^{10}$ (BBN), we obtain that $\Omega_{\rm x}\in [10^{-34}; 10^{-33}]$ at the $1\sigma$level, so the conventional BBN processes that occurred at a temperature of  $1 {\rm  Mev}$ are not spoiled  \cite{BBN}. 
\section{Summary}
We have studied an interacting dark matter and VVE scenario along with  decoupled baryons and photons components for a flat FRW universe; the  new nonlinear interaction allowed for the dark mix to interpolate between a warm dark matter regime at early times, after the initial  radiation regime,  and end with  a de Sitter phase.

The statistical analysis performed with the updated Hubble data [see Figs. (\ref{F1})-(\ref{F2}) and Table (\ref{I})] allowed us to constrain the behavior of dark energy in the recombination era and compare it with the latest bounds coming from the Planck+WP+high L data, SPT,  and ACT, among other experiments.  We have found that $\Omega_{\rm x}(z\simeq 10^{3}) \in [10^{-6}, 10^{-5}]$, so our estimations satisfied the stringent  bound reported by the Planck mission, $\Omega_{\rm ede} < 0.009$  at $95\%$ C.L. \cite{Planck2013} [see Table (\ref{II})] and agrees with  the small-scale CMB temperature measurement from the SPT  \cite{Reic} or with the upper limit set by WMAP7+SPT+BAO+SNe data \cite{Hou}. Further, the value $\Omega_{\rm x}(z\simeq 10^{3})\simeq 10^{-6}$ obtained here  will be consistent with the future constraints achievable by the Euclid and CMBPol experiments, \cite{Euc}, \cite{EDE2}. We also showed that dark energy around $z=10^{10}$ (BBN) fulfills the strong upper limit $\Omega_{\rm x}(z\simeq 10^{10})<0.04$  at the $1\sigma$ level \cite{BBN},  so the standard  BBN processes and  the well-measured abundance of light elements are not disturbed.
\acknowledgments
The authors thank the anonymous referee for comments that greatly improved the clarity of the manuscript.
L.P.C thanks  U.B.A under Project No. 20020100100147 and CONICET under Project PIP No. 114-200801-00328.  M.G.R and  I.S.G are supported by CONICET.



\end{document}